\documentclass[10.75pt,preprint]{article}
\usepackage[english]{babel}
\usepackage{slashed}
\usepackage{amsmath,amssymb,graphicx,calc,capt-of,ifthen,theorem,pifont,amsfonts}

\begin{document}

\title{Electroweak standard model with very special relativity}
\author{Jorge Alfaro, Pablo Gonz\'alez and Ricardo \'Avila.\\ \textit{Facultad de F\'isica, Pontificia Universidad Cat\'olica de Chile.}\\
\textit{Casilla 306, Santiago 22, Chile.}\\
jalfaro@uc.cl, pegonza2@uc.cl, raavila@uc.cl}

\maketitle

%%%%%%%%%%%%%%%%%%%%%%%%%%%%%%%%%%%%%Abstract%%%%%%%%%%%%%%%%%%%%%%%%%%%%%%%%%%%%%%%%%%%%%%%%%%%%%%%%%%%%%%%%%%%%%%%%%%%%%%%%%%%%%%%%%%%%%%%%

\section*{Abstract}
 The Very Special Relativity Electroweak Standard Model (VSR EW SM) is a theory with $SU(2)_L \times U(1)_R$ symmetry, with the same number of leptons and gauge fields as in the usual Weinberg-Salam (WS) model. No new particles are introduced. The model is renormalizable and unitarity is preserved. However, photons obtain mass and the massive bosons obtain different masses for different polarizations. Besides, neutrino masses are generated. A VSR invariant term will produce neutrino oscillations and new processes are allowed. In particular, we compute the rate of the decays $\mu \rightarrow e + \gamma$. All these processes, which are forbidden in the Electroweak Standard Model, put stringent bounds on the parameters of our model and measure the violation of Lorentz invariance. We investigate the canonical quantization of this non-local model. Second quantization is carried out obtaining a well defined particle concept. Additionally, we do a counting of the degrees of freedom associated to the gauge bosons involved in this work, after Spontaneous Symmetry Breaking has been realized. Violations of Lorentz invariance have been predicted by several theories of Quantum Gravity \cite{amu}. It is a remarkable possibility that the low energy effects of Lorentz violation induced by Quantum Gravity could be contained in the non-local terms of the VSR EW SM.

%%%%%%%%%%%%%%%%%%%%%%%%%%%%%%%%%%%%%Introduction%%%%%%%%%%%%%%%%%%%%%%%%%%%%%%%%%%%%%%%%%%%%%%%%%%%%%%%%%%%%%%%%%%%%%%%%%%%%%%%%%%%%%%%%%%%%%%%%

\section*{Introduction}

The $SU(2)_{L} \times U(1)_{R}$ gauge theory of weak and electromagnetic interactions known as the Electroweak Standard Model or Weinberg-Salam model(SM) is one of the most successful theories of Elementary Particle Physics. It permits to describe in detail an enormous amount of experimental data. Moreover, precision tests at the LHC, have verified both the particle contain, the gauge couplings as well as the mechanism of spontaneous symmetry breaking of the SM. The discovery of the Higgs particle with the SM properties at the LHC has completed the picture, leaving a very restrictive range of parameters to be explained by new Physics \cite{w2}. Therefore, any modification to the SM structure must be very subtle. Nonetheless, SM as it is cannot be the ultimate theory of nature. It does not incorporate the observed fact that the neutrinos have mass and does not incorporate gravity \cite{Langacker}.\\

The main problem of the Weinberg-Salam model  is the observation of neutrino flavor oscillations, which imply that neutrinos are massive\footnote{Although, it was expected that the neutrinos have mass since the 80's\cite{Perkins}}. The SM does not provide an explanation for this fact, since neutrinos are massless in it. If Lorentz symmetry is exact, additional massive particles must be postulated as in the popular seesaw mechanism \cite{mohapatra}. These remarks suggest that the subtle modification to the SM must be focused in the generation of neutrino mass and neutrino oscillations, preserving both the symmetries as well as the particles of the SM.\\

One possibility to include neutrino masses is to introduce a breaking of Lorentz symmetry, through constant background fields causing deviations of Lorentz symmetry \cite{Kostelecky}, but such proposals have as a consequence that the dispersion relation for light is modified.\\

A more conservative alternative would be to keep the essential features of special relativity, like the constancy of the velocity of light, but leave aside rotation symmetry with a subgroup of Lorentz. Such subgroups were identified and used to built what is called Very Special Relativity (VSR) \cite{CG1}. One of its main features is that the inclusion of P, T or CP symmetries enlarges VSR to the full Lorentz group. The most interesting of these subgroups are SIM(2) and HOM(2). These subgroups do not have invariant tensor fields besides the ones that are invariant under the whole Lorentz group, therefore the dispersion relations, time delay and all classical tests of SR are valid too. However, a non local term is necessary to formulate VSR \cite{CG2}.\\

VSR admits the generation of a neutrino mass without lepton number violation and without sterile neutrinos. Following this line of thought, in this paper, we study a modification of the Electroweak Standard Model using as the symmetry of nature VSR. One advantage of this model is that we do not necessitate to include more particles that the currently known. The VSR EW SM is a simple theory with $SU(2)_L \times U(1)_R$ symmetry, with the same number of leptons and gauge fields as in the usual Electroweak model. But now, it is possible to introduce new mass terms that violate Lorentz invariance. These terms are non-local and relevant at low energies and are able to describe in a straightforward manner the observed neutrino oscillations.\\

The gauge theory formalism necessary to implement the VSR EW SM has been recently developed in \cite{ar1}. The model is renormalizable and unitarity is preserved.\\

In the VSR EW SM new processes are allowed, which are consistent with the available data. Neutrino oscillations have the same form as in a Lorentz invariant theory. We also compute the decay rate of $\mu ->e+ \gamma$. All these processes, which are forbidden in the SM, put stringent bounds on the parameters of our model and measure the violation of Lorentz invariance.\\

Violations of Lorentz invariance have been predicted by several theories of Quantum Gravity \cite{amu}. It is an enticing possibility that the low energy
effects of Lorentz violation induced by Quantum Gravity are embodied in the non-local terms of the VSR EW SM.\\

We have organized the paper as follows: In section 1, we review the formulation of Yang Mills fields in VSR. In section 2, we define the VSR EW SM gauge bosons, using the formalism of section 1. Besides, the VSR Weinberg-Salam model is defined in its various components: gauge, leptons, scalars and interactions. Section 3 contains the description of Spontaneous Symmetry Breaking. Masses for gauge fields and leptons are computed. Section 4 contains the dispersion relations for the electron (muon,tau) for the general case $m_L\neq m_R$. Section 5 study a novel oscillation: Electron Spin oscillation due to VSR. Bounds on some VSR parameters are proposed. In Section 6 we write the leptons gauge bosons interactions and study the terms which are responsible of neutrino oscillations in the model. In section 7, we compute the decay rate for the process $X->Y+\gamma$ (flavor changing). Comparing with the best experimental bounds available today, we get more restrictions on the VSR parameters. Finally, Section 8 contains the canonical quantization of the (nonlocal) model. Finally, we have the conclusions.\\

Additionally, in Appendix A, we study the gauge bosons equation of motion and do the counting of degrees of freedom for the massive as well as the massless case. Appendix B has the solutions of the VSR Dirac equation for $m_L\neq m_R$ whereas in Appendix C we obtain the solution of the VSR Dirac equation for the particular but phenomenologically important case $m_L=m_R=m$.\\

Next, we use the results of {\cite{ar1}} to build the VSR EW SM based on $SU(2)_L \times U(1)_R$ group and the SM particle representations.\\

%%%%%%%%%%%%%%%%%%%%%%%%%%%%%%%%%%%%%Non Abelian Gauge Fields%%%%%%%%%%%%%%%%%%%%%%%%%%%%%%%%%%%%%%%%%%%%%%%%%%%%%%%%%%%%%%%%%%%%%%%%%%%%%%%%%%%%%%%%%%%%%%%%

\section{Non Abelian Gauge Fields}
\label{Chap: Non Abelian Gauge Fields.}

In this section we review the results of {\cite{ar1}}. We consider a scalar field transforming under a non-Abelian gauge transformation with infinitesimal parameter $\Lambda$:

\begin{eqnarray}
  \delta \phi =i \Lambda \phi.
\end{eqnarray}

We define the covariant derivative by:

\begin{eqnarray}
\label{cdna}
D_{\mu} \phi = \partial_{\mu} \phi - iA_{\mu} \phi + \frac{i}{2} m^{2}n_{\mu} \left((n \cdot \partial )^{-2} (n \cdot A)\right) \phi,
\end{eqnarray}

where $m$ is a constant with dimensions of mass. It measures the departure from Lorentz invariance, since in the term containing it in (\ref{cdna}) a fixed null vector $n_{\mu}$ appears. This non-local term is invariant under Sim(2) and Hom(2) because the transformations of these subgroups of the Lorentz group change $n_{\mu}$ at most by a multiplicative constant factor, which is canceled out by the change of $n_{\mu}$ in the denominator. Now, we impose as usual that:

\begin{eqnarray}
\label{cdd}
\delta \left(D_{\mu} \phi\right) = i \Lambda D_{\mu} \phi.
\end{eqnarray}

Then the gauge transformation for the gauge field is:

\begin{eqnarray}
\delta_{\Lambda} A_{\mu} &=& \partial_{\mu} \Lambda - i[A_{\mu} , \Lambda ] - \frac{i m^2}{2} n_{\mu}\left[\Lambda, \left((n \cdot \partial)^{-2}(n\cdot A)\right)\right] + \frac{m^2}{2}n_{\mu}\left((n\cdot\partial)^{-1} \Lambda\right) \nonumber \\
&& - \frac{i m^2}{2}n_{\mu}\left((n \cdot \partial )^{-2}(n\cdot[A, \Lambda])\right).
\end{eqnarray}

We have also checked the closure of the algebra:

\begin{eqnarray}
\left[\delta_{\Lambda_{1}},\delta_{\Lambda_{2}}\right]A_{\mu} = - i\delta_{\left[\Lambda_{1},\Lambda_{2}\right]} A_{\mu}.
\end{eqnarray}

The commutator of two covariant derivatives defines $F_{\mu \nu}$, the $A_{\mu}$ Field Strength:

\begin{eqnarray}
\label{fs1}
[D_{\mu},D_{\nu}] \phi = - i F_{\mu \nu} \phi,
\end{eqnarray}

so we get:

\begin{eqnarray}
\label{F}
F_{\mu \nu} &=& A_{\nu, \mu} - A_{\mu, \nu} - i [A_{\mu} ,A_{\nu} ] - \frac{m^2}{2} n_{\nu}\left((n\cdot\partial)^{-2} (n\cdot A_{,\mu})\right) + \frac{m^2}{2} n_{\mu}\left((n \cdot \partial)^{-2} (n \cdot A_{,\nu})\right) \nonumber \\
&& + \frac{im^2}{2}\left[\left((n \cdot \partial )^{-2} (n\cdot A)\right),(n_{\mu} A_{\nu} - n_{\nu} A_{\mu})\right].
\end{eqnarray}

It is hermitian if $A_{\mu}$ is hermitian:

\begin{eqnarray}
[D'_{\mu} ,D'_{\nu} ] \phi' = U [D_{\mu} ,D_{\nu} ] \phi = U(-iF_{\mu \nu} )\phi = (-iF'_{\mu \nu} )U \phi \textrm{ with: } U = e^{i \Lambda},
\end{eqnarray}

then we find:

\begin{eqnarray}
F'_{\mu \nu} = UF_{\mu \nu} U^{-1}.
\end{eqnarray}

It is not difficult to see that a redefinition given by $A_{\mu} \rightarrow A_{\mu} - \frac{1}{2} m^{2}n_{\mu} \left((n \cdot \partial )^{-2} (n \cdot A)\right)$ eliminate the modification by the $m$ factor. This means that a modification in the ordinary covariant derivative given by (\ref{cdna}) do not affect the observables. Then, we will use $m = 0$ from now on. However, VSR allow us to define a new invariant mass term for gauge fields using a new Field Strength:

\begin{eqnarray}
\label{F new}
\tilde{F}_{\mu \nu} &=& F_{\mu \nu} + \frac{m_A^2}{2}\left(n_{\nu}\frac{1}{(n \cdot D)^2}(n^{\alpha}F_{\mu \alpha}) - n_{\mu}\frac{1}{(n\cdot D)^2}(n^{\alpha}F_{\nu \alpha})\right).
\end{eqnarray}

We will develop the effect of this element in the next section.\\

Finally, we define the wiggle covariant derivative of the field $\phi$ by:

\begin{eqnarray}
\label{wcdna}
\tilde{D}_{\mu} \phi = D_{\mu} \phi + \frac{1}{2}\frac{m^{2}_{\phi}}{n\cdot D}n_{\mu}\phi,
\end{eqnarray}

where $m^{2}_{\phi}$ is a new VSR parameter. Using $ \tilde{D}_{\mu}$ we can introduce different VSR masses for the various matter fields in a covariant manner.\\

%%%%%%%%%%%%%%%%%%%%%%%%%%%%%%%%%%%%%Very Special Relativity Standard Model%%%%%%%%%%%%%%%%%%%%%%%%%%%%%%%%%%%%%%%%%%%%%%%%%%%%%%%%%%%%%%%%%%%%%%%%%%%%%%%%%%%%

\section{Very Special Relativity Electroweak Standard Model}
\label{Chap: Very Special Relativity Electroweak Standard Model.}

In the Electroweak model, we have a symmetry given by $SU(2)_L \times U(1)_R$, so a generic field, $\psi$, will transform like:

\begin{eqnarray}
\label{trans psi}
\delta \psi = i(\mathbf{\Lambda} + \Theta)\psi,
\end{eqnarray}

where $\mathbf{\Lambda}$ and $\Theta$ are transformation parameters under SU(2) and U(1) respectively. To define the covariant derivative, we must impose:

\begin{eqnarray}
\label{trans Dpsi}
\delta (D_{\mu}\psi) = i(\mathbf{\Lambda} + \Theta)D_{\mu}\psi.
\end{eqnarray}

We saw in the last section that for VSR the covariant derivative is not modified. Then:

\begin{eqnarray}
\label{Der Cov}
D_{\mu} \psi = \partial_{\mu} \psi - ig'\mathbf{B}_{\mu}\psi - ig\mathbf{A}_{\mu}\psi
\end{eqnarray}

where $\mathbf{A}_{\mu} = \frac{\tau_i}{2} A^i_{\mu}$ and $\mathbf{B}_{\mu} = \frac{Y}{2} B_{\mu}$. Besides, we have:

\begin{eqnarray}
\label{trans A 1}
\delta \mathbf{A}_{\mu} &=& \frac{1}{g}\partial_{\mu} \mathbf{\Lambda} - i\left[\mathbf{A}_{\mu},\mathbf{\Lambda}\right] \\
\label{trans B 1}
\delta \mathbf{B}_{\mu} &=& \frac{1}{g'}\partial_{\mu} \Theta
\end{eqnarray}

or taken the Lie algebra components:

\begin{eqnarray}
\label{trans A 2}
\delta A^i_{\mu} &=& \frac{1}{g}\partial_{\mu} \epsilon^i + \varepsilon^i_{~jk}A^j_{\mu}\epsilon^k \\
\label{trans B 2}
\delta B_{\mu} &=& \frac{1}{g'}\partial_{\mu} \epsilon^0,
\end{eqnarray}

where $\varepsilon^i_{~jk}$ is the Levi-Civita symbol and we used that $\mathbf{\Lambda} = \frac{\tau_i}{2}\epsilon^i$ and $\Theta = \frac{Y}{2}\epsilon^0$. The ordinary Field Strength on VSR for both gauge fields are given by:

\begin{eqnarray}
\label{D com}
[D_{\mu},D_{\nu}]\psi = - i \left(g\frac{\tau_i}{2}F^i_{\mu \nu} + g'\frac{Y}{2}B_{\mu \nu}\right) \psi,
\end{eqnarray}

such that:

\begin{eqnarray}
\label{FS F}
F^i_{\mu \nu} &=& \partial_{\mu} A^i_{\nu} - \partial_{\nu} A^i_{\mu} + g\varepsilon^i_{~jk}A^j_{\mu}A^k_{\nu} \\
\label{FS B}
B_{\mu \nu} &=& \partial_{\mu} B_{\nu} - \partial_{\nu} B_{\mu}.
\end{eqnarray}

But, using (\ref{F new}), we can define:

\begin{eqnarray}
\label{F tild}
\tilde{F}^i_{\mu \nu} &=& F^i_{\mu \nu} + \frac{m_A^2}{2}\left(n_{\nu}\frac{1}{(n \cdot D)^2}(n^{\alpha}F^i_{\mu \alpha}) - n_{\mu}\frac{1}{(n\cdot D)^2}(n^{\alpha}F^i_{\nu \alpha})\right) \\
\label{B tild}
\tilde{B}_{\mu \nu} &=& B_{\mu \nu} + \frac{m_B^2}{2}\left(n_{\nu}\frac{1}{(n \cdot \partial)^2}(n^{\alpha}B_{\mu \alpha}) - n_{\mu}\frac{1}{(n\cdot \partial)^2}(n^{\alpha}B_{\nu \alpha})\right)
\end{eqnarray}

Now, we have all the elements to build the Weinberg-Salam model on VSR. For this we need the gauge fields $B_{\mu}$ and $A^i_{\mu}$, three families of leptons and a scalar field to implement the Higgs mechanism. Then we have:\\

\textbf{I) Gauge Lagrangian:} Two kind of gauge fields, $B_{\mu}$ and $A^i_{\mu}$. To write the lagrangian, we use the modified Fields Strength given by (\ref{F tild}) and (\ref{B tild}). Then:

\begin{eqnarray}
\label{L gauge 0}
\mathcal{L}_{gauge} =  - \frac{1}{4}\tilde{F}^i_{\mu \nu}\tilde{F}_i^{\mu \nu} - \frac{1}{4}\tilde{B}_{\mu \nu}\tilde{B}^{\mu \nu}.
\end{eqnarray}

We can prove that:

\begin{eqnarray}
\tilde{F}^i_{\mu \nu}\tilde{F}_i^{\mu \nu} &=& F^i_{\mu \nu}F_i^{\mu \nu} + 2m_A^2(n^{\alpha}F^i_{\mu \alpha})(n \cdot D)^{-2}(n_{\beta}F_i^{\mu \beta}) \\
\tilde{B}_{\mu \nu}\tilde{B}^{\mu \nu} &=& B_{\mu \nu}B^{\mu \nu} + 2m_B^2(n^{\alpha}B_{\mu \alpha})(n \cdot \partial)^{-2}(n_{\beta}B^{\mu \beta}).
\end{eqnarray}

Therefore the lagrangian is now:

\begin{eqnarray}
\label{L gauge}
\mathcal{L}_{gauge} =  - \frac{1}{4}F^i_{\mu \nu}F_i^{\mu \nu} - \frac{m_A^2}{2}(n^{\alpha}F^i_{\mu \alpha})(n \cdot D)^{-2}(n_{\beta}F_i^{\mu \beta}) - \frac{1}{4}B_{\mu \nu}B^{\mu \nu} - \frac{m_B^2}{2}(n^{\alpha}B_{\mu \alpha})(n \cdot \partial)^{-2}(n_{\beta}B^{\mu \beta}).
\end{eqnarray}

From this lagrangian we can see that the equation of motion of $B_{\mu}$ is:

\begin{eqnarray}
\partial_{\nu}B^{\mu \nu} - m_B^2n^{\mu}(n \cdot \partial)^{-2}\partial_{\alpha}(n_{\beta}B^{\alpha \beta}) + m_B^2(n \cdot \partial)^{-1}(n_{\beta}B^{\mu \beta}) = 0.
\end{eqnarray}

Now, if we contract this equation with $n_{\mu}$, we obtain $\partial_{\nu}(n_{\mu}B^{\mu \nu}) = 0$, so:

\begin{eqnarray}
\label{Id 1}
\partial{\nu}B^{\mu \nu} + m_B^2(n \cdot \partial)^{-1}(n_{\beta}B^{\mu \beta}) &=& 0 \nonumber \\
\rightarrow \partial^2 B_{\mu} - \partial_{\mu}\partial_{\nu} B^{\nu} + m_B^2 B_{\mu} - m_B^2(n \cdot \partial)^{-1}\partial_{\mu}(n \cdot B) &=& 0
\end{eqnarray}

and:

\begin{eqnarray}
\label{Id 2}
\partial_{\nu}(n_{\mu}B^{\mu \nu})  &=& 0 \nonumber \\
\rightarrow \partial^2 (n \cdot B) - (n \cdot \partial)\partial_{\nu} B^{\nu} &=& 0.
\end{eqnarray}

On the other side, we need to fix the gauge freedom. We can use the Lorentz gauge plus a VSR additional restriction given by:

\begin{eqnarray}
\partial_{\mu} B^{\mu} = 0 \\
n_{\mu} B^{\mu} = 0.
\end{eqnarray}

If we use it in (\ref{Id 1}) and (\ref{Id 2}), we obtain:

\begin{eqnarray}
\partial^2 B^{\mu} + m_B^2 B_{\mu} = 0.
\end{eqnarray}

From this equation we can see that $B^{\mu}$ have a mass $m_B$. A similar result we can obtain from the free equation of motion of $A^i_{\mu}$, where the mass is $m_A$. Therefore, the lagrangian (\ref{L gauge}) describes massive gauge fields, but we will see that it preserves two degrees of freedom (\textbf{See Appendix A}).\\

In \textbf{Section \ref{Sec: Gauge Fields}}, we will study the free dynamic using (\ref{L gauge}) after the Spontaneous Symmetry Breaking is realized.\\

\textbf{II) Leptonic Lagrangian:} Three $SU(2)$ doublets $L_a = \left(
                               \begin{array}{c}
                                   \nu^0_{aL} \\
                                     e^0_{aL} \\
                               \end{array}
                            \right)$, where $\nu^0_{aL} = \frac{1}{2}(1-\gamma_5)\nu^0_{a}$ and $e^0_{aL} = \frac{1}{2}(1-\gamma_5)e^0_{a}$, and three $SU(2)$ singlet $R_a = e^0_{aR} = \frac{1}{2}(1+\gamma_5)e^0_{n}$. As is usual, we make the supposition that there is no right-handed neutrino. The index $a$ represent the different families and the index $0$ say that the fermionic fields are the physical fields before breaking the symmetry of the vacuum. The lagrangian is:
\begin{eqnarray}
\label{L lepton}
\mathcal{L}_{lepton} &=& \bar{L}_{b}i\gamma^{\mu}\left[\tilde{D}^{(L)}_{\mu}\right]^{b a}L_{a} + \bar{R}_{b}i\gamma^{\mu}\left[\tilde{D}^{(R)}_{\mu}\right]^{b a}R_{a} \nonumber \\
&=& \bar{L}_{b}i\gamma^{\mu}\left(\delta^{b a}D_{\mu} + \frac{1}{2}[m_L^2]^{b a} n_{\mu} (n^{\alpha}D_{\alpha})^{-1}\right)L_{a} + \bar{R}_{b}i\gamma^{\mu}\left(\delta^{b a}D_{\mu} + \frac{1}{2}[m_R^2]^{b a} n_{\mu} (n^{\alpha}D_{\alpha})^{-1}\right)R_{a} \nonumber \\
&=& i\bar{L}^a\slashed{D}L_a + \frac{i}{2}\bar{L}_b\slashed{n}[m_L^2]^{b a} (n^{\alpha}D_{\alpha})^{-1}L_a + i\bar{R}^a\slashed{D}R_a + \frac{i}{2}\bar{R}_b\slashed{n}[m_R^2]^{b a} (n^{\alpha}D_{\alpha})^{-1}R_a,
\end{eqnarray}

where $m_L^2$ and $m_R^2$ are hermitian matrices in family indices, $(ba)$, \textit{and they could depend on $\gamma^5$}. The doublets have a hypercharge $Y = -1$ and the singlets have $Y = -2$. So, using (\ref{Der Cov}), we have:

\begin{eqnarray}
\label{DL DR}
D_{\mu}L_a &=& \left(\partial_{\mu} + \frac{ig'}{2}B_{\mu} - ig\frac{\tau_i}{2}A^i_{\mu}\right) L_a \nonumber \\
D_{\mu}R_a &=& \left(\partial_{\mu} + ig'B_{\mu}\right) R_a.
\end{eqnarray}

We will see that $m_L^2$ is the mass matrix of neutrinos, that generate the oscillation between the different families.\\

\textbf{III) Scalar Lagrangian:} A complex doublet scalar field $\phi = \left(
                                                 \begin{array}{c}
                                                    \phi^{+} \\
                                                    \phi^0 \\
                                                 \end{array}
                                               \right)$ with a lagrangian given by:

\begin{eqnarray}
\mathcal{L}_{scalar} &=& (\tilde{D}_{(\phi)}^{\mu}\phi)^{\dag}(\tilde{D}^{(\phi)}_{\mu}\phi) - V\left(\phi^{\dag}\phi\right) \nonumber \\
&=& (D^{\mu}\phi)^{\dag}(D_{\mu}\phi) - m_{\phi}^2\phi^{\dag}\phi - V\left(\phi^{\dag}\phi\right), \nonumber
\end{eqnarray}

with:

\begin{eqnarray}
\label{V}
V\left(\phi^{\dag}\phi\right) = \mu^2\phi^{\dag}\phi + \lambda(\phi^{\dag}\phi)^2.
\end{eqnarray}

We notice that the term proportional to $m_{\phi}^2$ can be absorbed redefining $\mu^2 + m_{\phi}^2 \rightarrow \mu^2$. Therefore, our scalar lagrangian is reduced to:

\begin{eqnarray}
\label{L scalar}
\mathcal{L}_{scalar} &=& (D^{\mu}\phi)^{\dag}(D_{\mu}\phi) - \mu^2(\phi^{\dag}\phi) - \lambda (\phi^{\dag}\phi)^2.
\end{eqnarray}

The hypercharge of $\phi$ is $Y = 1$, so:

\begin{eqnarray}
\label{D phi}
D_{\mu}\phi = \left(\partial_{\mu} - \frac{ig'}{2}B_{\mu} - ig\frac{\tau_i}{2}A^i_{\mu}\right) \phi.
\end{eqnarray}

From (\ref{D phi}), we will obtain masses of each fields after Spontaneous Symmetry Breaking. The dynamic of the Higgs is not important, for the moment, so we will focus in the gauge and lepton fields.\\

\textbf{IV) Interaction Lagrangian:} Besides, we have an interaction lagrangian:

\begin{eqnarray}
\label{L interaction}
\mathcal{L}_{int} = - \left[\Gamma\right]^{b a} \bar{L}_b\phi R_a - \left[\Gamma^{\dag}\right]^{b a}\bar{R}_b \phi^{\dagger} L_a,
\end{eqnarray}

where $\Gamma$ is a matrix associated to the Yukawa interaction. Therefore, the final lagrangian is given by:

\begin{eqnarray}
\mathcal{L} = \mathcal{L}_{gauge} + \mathcal{L}_{lepton} + \mathcal{L}_{scalar} + \mathcal{L}_{int}.
\end{eqnarray}

Now, we can proceed to break the symmetry, using the Higgs mechanism.\\

%%%%%%%%%%%%%%%%%%%%%%%%%%%%%%%%%%%%%%%%%Spontaneous Symmetry Breaking%%%%%%%%%%%%%%%%%%%%%%%%%%%%%%%%%%%%%%%%%%%%%%%%%%%%%%%%%%%%%%%%%%%%%%%%%%%%%%%%%%%%%%%%%%

\section{Spontaneous Symmetry Breaking}
\label{Chap: Spontaneous Symmetry Breaking.}

To break the symmetry, we need to find the vacuum of the system. For this we search for a solution to $\frac{\partial V}{\partial \phi} = 0$. Seeing (\ref{V}), it can be noticed that this occurs for $\langle \phi \rangle = 0$ or $\langle \phi \rangle = \frac{1}{\sqrt{2}}\left(
                                                      \begin{array}{c}
                                                        0 \\
                                                        v \\
                                                      \end{array}
                                                    \right)$, where $v= \sqrt{\frac{-\mu^2}{\lambda}}$. It is useful to work in the unitary gauge. In this gauge, the Goldstone bosons are
removed from the lagrangian doing a gauge transformation. After the gauge transformation, we can use
$\phi = \left( \begin{array}{c}
  0\\
  \frac{v+H}{\sqrt{2}}
\end{array} \right)$, where $H$ is the Higgs boson. From the non-zero value for the vacuum, we have new quadratic term in the fields, so they will obtain
an additional contribution to the  mass. To compute the mass, we will study the free part in the lagrangian for each field.\\

Notice that:

\begin{eqnarray}
\left( T_{L}^{3} + \frac{Y}{2} \right)
\left( \begin{array}{c}
    0\\
    v
\end{array} \right)
=
\left( \begin{array}{c}
    0\\
    0
\end{array} \right)
= Q <\phi>
\end{eqnarray}

So the operator $Q=T_{L}^{3} + \frac{Y}{2}$ is still a symmetry of the vacuum.
The associated gauge field will remain with the VSR mass only.\\

\subsection{Gauge Fields:}
\label{Sec: Gauge Fields}

After Spontaneous Symmetry Breaking, we have contributions to quadratic term in the gauge fields from (\ref{L gauge}) and (\ref{L scalar}). Then:

\begin{eqnarray}
\label{L gauge free 0}
\mathcal{L}_{\textrm{gauge free}} &=& \left(- \frac{1}{4}F^i_{\mu \nu}F_i^{\mu \nu} - \frac{m_A^2}{2}(n^{\alpha}F^i_{\mu \alpha})(n \cdot D)^{-2}(n_{\beta}F_i^{\mu \beta})\right)_{(2)} - \frac{1}{4}B_{\mu \nu}B^{\mu \nu} - \frac{m_B^2}{2}(n^{\alpha}B_{\mu \alpha})(n \cdot \partial)^{-2}(n_{\beta}B^{\mu \beta}) \nonumber \\
&& + \frac{v^2}{8}
\left(
  \begin{array}{cc}
    0 & 1 \\
  \end{array}
\right)
\left(g'B_{\mu} + g\tau_iA^i_{\mu}\right)^{\dag}\left(g'B_{\mu} + g\tau_jA^j_{\mu}\right)
\left(
  \begin{array}{c}
    0 \\
    1 \\
  \end{array}
\right),
\end{eqnarray}

where the Pauli matrices are:

\begin{eqnarray}
\label{Pauli}
\tau_1 =
\left(
  \begin{array}{cc}
    0 & 1 \\
    1 & 0 \\
  \end{array}
\right) \textrm{, }
\tau_2 =
\left(
  \begin{array}{cc}
    0 & -i \\
    i &  0 \\
  \end{array}
\right) \textrm{, }
\tau_3 =
\left(
  \begin{array}{cc}
    1 &  0 \\
    0 & -1 \\
  \end{array}
\right)
\end{eqnarray}

and the subindex $(2)$ means that we keep up to quadratic terms in the field. Evaluating (\ref{FS F}), (\ref{FS B}) and (\ref{Pauli}) in (\ref{L gauge free
0}), we obtain:

\begin{eqnarray}
\label{L gauge free 1}
\mathcal{L}_{\textrm{gauge free}} &=& \frac{1}{2}A^i_{\mu}\left((\partial^2 + m_A^2)\delta^{\mu}_{\nu}  - (\partial^{\mu} + m_A^2(n \cdot \partial)^{-1}n^{\mu})(\partial_{\nu} + m_A^2(n \cdot \partial)^{-1}n_{\nu}) + m_A^2n^{\mu}n_{\nu}(n \cdot \partial)^{-2}(\partial^2 + m_A^2)\right)A_i^{\nu} \nonumber \\
&& + \frac{1}{2}B_{\mu}\left((\partial^2 + m_B^2)\delta^{\mu}_{\nu}  - (\partial^{\mu} + m_B^2(n \cdot \partial)^{-1}n^{\mu})(\partial_{\nu} + m_B^2(n \cdot \partial)^{-1}n_{\nu}) + m_B^2n^{\mu}n_{\nu}(n \cdot \partial)^{-2}(\partial^2 + m_B^2)\right)B^{\nu} \nonumber \\
&& + \frac{v^2 g^2}{8}\left(A_1^{\mu}A^1_{\mu} + A_2^{\mu}A^2_{\mu}\right) + \frac{v^2}{8}\left(g'B_{\mu}-gA^3_{\mu}\right)\left(g'B^{\mu}-gA_3^{\mu}\right)
\end{eqnarray}

Now, to diagonalize this Lagrangian, we study the case $m_A = m_B \equiv m_G$ and:

\begin{eqnarray}
\label{new gauge}
A^1_{\mu} &=&  \frac{1}{\sqrt{2}}\left(W^+_{\mu} + W^-_{\mu}\right) \nonumber \\
A^2_{\mu} &=&  \frac{i}{\sqrt{2}}\left(W^+_{\mu} - W^-_{\mu}\right) \nonumber \\
A^3_{\mu} &=& \frac{g'A_{\mu} -g Z_{\mu}}{\sqrt{g^2+g'^2}} \nonumber \\
B_{\mu} &=& \frac{g A_{\mu} + g'Z_{\mu}}{\sqrt{g^2+g'^2}}.
\end{eqnarray}

Then:

\begin{eqnarray}
\label{L gauge free}
\mathcal{L}_{\textrm{gauge free}} &=& W^{-}_{\mu}\left(\left(\partial^2 + m_G^2 + \frac{v^2g^2}{4}\right)\delta^{\mu}_{\nu} - (\partial^{\mu} + m_G^2(n \cdot \partial)^{-1}n^{\mu})(\partial_{\nu} + m_G^2(n \cdot \partial)^{-1}n_{\nu})\right. \nonumber \\
&&\left. + m_G^2n^{\mu}n_{\nu}(n \cdot \partial)^{-2}(\partial^2 + m_G^2)\right)W_{+}^{\nu} \nonumber \\
&& + \frac{1}{2}Z_{\mu}\left(\left(\partial^2 + m_G^2 + \frac{v^2(g^2+g'^2)}{4}\right)\delta^{\mu}_{\nu}  - (\partial^{\mu} + m_G^2(n \cdot \partial)^{-1}n^{\mu})(\partial_{\nu} + m_G^2(n \cdot \partial)^{-1}n_{\nu}) \right. \nonumber \\
&&\left. + m_G^2n^{\mu}n_{\nu}(n \cdot \partial)^{-2}(\partial^2 + m_G^2)\right)Z^{\nu} \\
&& + \frac{1}{2}A_{\mu}\left((\partial^2 + m_G^2)\delta^{\mu}_{\nu}  - (\partial^{\mu} + m_G^2(n \cdot \partial)^{-1}n^{\mu})(\partial_{\nu} + m_G^2(n \cdot \partial)^{-1}n_{\nu}) \right. \nonumber \\
&&\left. + m_G^2n^{\mu}n_{\nu}(n \cdot \partial)^{-2}(\partial^2 + m_G^2)\right)A^{\nu}. \nonumber
\end{eqnarray}

From this Lagrangian, we can find the free equations of motion. They are:

\begin{eqnarray}
\label{EQ W}
\left(\partial^2 + m_G^2 + \frac{v^2g^2}{4}\right)W^{\pm}_{\mu} - (\partial_{\mu} + m_G^2(n \cdot \partial)^{-1}n_{\mu})\left((\partial \cdot W^{\pm}) + m_G^2(n \cdot \partial)^{-1}(n \cdot W^{\pm})\right) && \nonumber \\
+ m_G^2n_{\mu}(n \cdot \partial)^{-2}(\partial^2 + m_G^2)(n \cdot W^{\pm}) &=& 0 \\
\label{EQ Z}
\left(\partial^2 + m_G^2 + \frac{v^2(g^2+g'^2)}{4}\right)Z_{\mu} - (\partial_{\mu} + m_G^2(n \cdot \partial)^{-1}n_{\mu})\left((\partial \cdot Z) + m_G^2(n \cdot \partial)^{-1}(n \cdot Z)\right) && \nonumber \\
+ m_G^2n_{\mu}(n \cdot \partial)^{-2}(\partial^2 + m_G^2)(n \cdot Z) &=& 0 \\
\label{EQ A}
\left(\partial^2 + m_G^2\right)A_{\mu} - (\partial_{\mu} + m_G^2(n \cdot \partial)^{-1}n_{\mu})\left((\partial \cdot A) + m_G^2(n \cdot \partial)^{-1}(n \cdot A)\right) && \nonumber \\
+ m_G^2n_{\mu}(n \cdot \partial)^{-2}(\partial^2 + m_G^2)(n \cdot A) &=& 0.
\end{eqnarray}

Notice that all gauge field equations are like:

\begin{eqnarray}
\label{eqn}
\left(\partial^2 + M^2\right) V_{\mu} - \left(\partial_{\mu} + m_G^2(n \cdot \partial)^{-1} n_{\mu}\right)\left((\partial \cdot V) + m_G^2(n \cdot \partial )^{-1} (n \cdot V)\right) && \nonumber \\
+ m_G^2n_{\mu}(n \cdot \partial )^{-2}\left(\partial^2 + m_G^2\right)(n \cdot V) &=& 0,
\end{eqnarray}

where $M$ is a mass term. Using the result in (\textbf{Appendix A}), we obtain:\\

$W^{\pm}_{\mu}$ has three degrees of freedom. The mass is $M_{W} = \sqrt{\frac{v^2g^2}{4} + m_G^2}$ for  polarizations perpendicular to $n_{\mu}$ and $M_{W} = \frac{vg}{2}$ for the longitudinal polarization.\\

$Z_{\mu}$ has three degrees of freedom. The mass is $M_{Z} = \sqrt{\frac{v^2(g^2+g'^2)}{4} + m_G^2}$ for  polarizations perpendicular to $n_{\mu}$ and $M_{Z} = \frac{v\sqrt{g^2+g'^2}}{2}$ for the longitudinal polarization.\\

$A_{\mu}$ has two degrees of freedom and the mass is only $M_{A} = m_G$.\\

One thing that we must consider is the fact that the photon gains mass with VSR. Of course, this mass must be very tiny. Some widely accepted bounds for photon mass are:

\begin{itemize}
  \item Most accepted bound based in measuring the torque exerted on a magnetized ring caused by the galactic vector potential can be probed directly giving $m_G \leq 10^{-18}$ eV \cite{m phot 1}.
  \item Measures of the galactic magnetic field are only possible if the photon mass is zero, this has given a constraint of $m_G \leq 3 \times 10^{-27}$ eV \cite{m phot 2}.
\end{itemize}

On the other side, $W_{\mu}$ and $Z_{\mu}$ bosons  will exhibit different propagations for perpendicular and longitudinal polarizations respectively, just like birefringence, but the difference is extremely small since depends on the photon mass $m_G$. The bounds of photon mass give us a great idea how much similar are the masses of $W_{\mu}$ or $Z_{\mu}$ for different polarizations. It is certainly a prediction that should be investigated in appropriated experiments, for example at the LHC.\\

\subsection{Lepton Fields}

In order to see what happen to the leptons, we look at the diagonal (in flavor) part of $\mathcal{L}_{int}$. In particular, we
will study in more detail the electron family. $\mathcal{L}_{int}$ is now:

\begin{eqnarray}
\mathcal{L}_{int} = - \frac{G_ev}{\sqrt{2}} \left(\bar{e}_Re_L + \bar{e}_Le_R\right) + \textrm{ higher order terms,}
\end{eqnarray}

To determine the mass eigenstates we look at the equations of motion provided by the quadratic piece of the lagrangian. Introducing
$\psi =
\left(\begin{array}{c}
  e_{R}\\
  e_{L}
\end{array}\right)$, we get:

\begin{eqnarray}
\label{Eq psi}
\left(i\left(\slashed{\partial} + \frac{1}{2}\slashed{n}\bar{m}^2(n\cdot \partial)^{-1}\right) - \frac{G_ev}{\sqrt{2}}\right)\psi = 0,
\end{eqnarray}

where $\bar{m}^2 = m_R^2P_R + m_L^2P_L$, $P_L = \frac{1-\gamma^{5}}{2}$ and $P_R = \frac{1+\gamma^{5}}{2}$. Instead for the neutrino, we get:

\begin{eqnarray}
\label{Eq neutrino}
\left(i\left(\slashed{\partial} + \frac{m_L^2}{2}\slashed{n}(n\cdot \partial)^{-1}\right)\right)\nu_L = 0.
\end{eqnarray}

That is, the neutrino mass is $m_{\nu} = m_L$\footnote{The only pole the neutrino propagator has is at $p^{2} =m_L^{2}$. Please see Appendix D.}.\\

%%%%%%%%%%%%%%%%%%%%%%%%%%%%%%%%%%%%%%%%%Dispersion relations for m_{L} \neq m_{R}%%%%%%%%%%%%%%%%%%%%%%%%%%%%%%%%%%%%%%%%%%%%%%%%%%%%%%%%%%%%%%%%%%%%%%%%%%%%%

\section{Dispersion relations for $m_{L} \neq m_{R}$}
\label{Chap: Dispersion relations for mL mR.}

In this section, we write the solution of the VSR Dirac equation for the electron for the case $m_{L} \neq m_{R}$.\\

Particle:

\begin{eqnarray}
\label{Eq u}
\left(\slashed{p} - \frac{1}{2}\slashed{n}\bar{m}^2(n\cdot p)^{-1} - \frac{G_ev}{\sqrt{2}}\right)u_s = 0,
\end{eqnarray}

where:

\begin{eqnarray}
\label{u1}
u_1 &=& \frac{1}{2(n\cdot p)}\left(\slashed{p}+\frac{G_ev}{\sqrt{2}}\right)\slashed{n}\mathcal{U}_1 \textrm{, with: } p^2 = m_R^2 + \frac{G_e^2v^2}{2} \textrm{ and } p_0>0 \\
\label{u2}
u_2 &=& \frac{1}{2(n\cdot p)}\left(\slashed{p}+\frac{G_ev}{\sqrt{2}}\right)\slashed{n}\mathcal{U}_2 \textrm{, with: } p^2 = m_L^2 + \frac{G_e^2v^2}{2} \textrm{ and } p_0>0.
\end{eqnarray}

Antiparticle:

\begin{eqnarray}
\label{Eq v}
\left(\slashed{p} - \frac{1}{2}\slashed{n}\bar{m}^2(n\cdot p)^{-1} + \frac{G_ev}{\sqrt{2}}\right)v_s = 0,
\end{eqnarray}

where:

\begin{eqnarray}
\label{v1}
v_1 &=& \frac{1}{2(n\cdot p)}\left(\slashed{p}-\frac{G_ev}{\sqrt{2}}\right)\slashed{n}\mathcal{U}_1 \textrm{, with: } p^2 = m_R^2 + \frac{G_e^2v^2}{2} \textrm{ and } p_0>0 \\
\label{v2}
v_2 &=& \frac{1}{2(n\cdot p)}\left(\slashed{p}-\frac{G_ev}{\sqrt{2}}\right)\slashed{n}\mathcal{U}_2 \textrm{, with: } p^2 = m_L^2 + \frac{G_e^2v^2}{2} \textrm{ and } p_0>0.
\end{eqnarray}

$\mathcal{U}_1$ and $\mathcal{U}_2$ are constant spinors (See \textbf{Appendix B}). Therefore, if $m_{L} \neq m_{R}$, the electron, muon and tau are actually composed by two different particles with slightly different masses. Due to this, in the next section, we explore a novel electron oscillation and put
some plausible bounds on $|m_L - m_R|$.\\

%%%%%%%%%%%%%%%%%%%%%%%%%%%%%%%%%%%%%%%%%Electron spin precession%%%%%%%%%%%%%%%%%%%%%%%%%%%%%%%%%%%%%%%%%%%%%%%%%%%%%%%%%%%%%%%%%%%%%%%%%%%%%

\section{Electron spin precession}
\label{Chap: Electron spin precession.}

Consider an electron at rest with the spin up in the $z$ direction. Since $m_{L}$ and $m_{R}$ are expected to be much smaller than the contribution to the electron mass given by SSB, in a perturbative approach it makes sense to use the spinors that solve the Dirac equation with $m_{L} =m_{R} =0$ to describe the initial and final state of the electron. So, the probability to measure the spin down will be proportional to:

\begin{eqnarray}
P\left(\uparrow \rightarrow \downarrow\right) = 2R^{2}\left(1- \cos \left((E_{\uparrow} - E_{\downarrow})t\right)\right) = 4R^2\sin^{2}\left(\frac{(E_{\uparrow}-E_{\downarrow})t}{2}\right) \nonumber
\end{eqnarray}

for a certain constant $R$ and:

\begin{eqnarray}
E_{\uparrow} - E_{\downarrow} = \sqrt{m_{L}^{2}+M^{2}} - \sqrt{m_{R}^{2}  + M^{2}} \simeq M \left(1 + \frac{1}{2} \frac{m_{L}^{2}}{M^{2}} - 1 - \frac{1}{2}\frac{m_{R}^{2}}{M^{2}} \right) &=& \frac{\left(m_L^2 - m_R^2\right)}{2M} \nonumber \\
\frac{2\pi}{T} &=& \frac{\left|m_{L}^{2} - m_{R}^{2}\right|}{4M} \nonumber \\
\rightarrow T &=& \frac{8 \pi M}{\left|m_{L}^{2} - m_{R}^{2}\right|}.
\end{eqnarray}

To put some bound on this effect, we can imagine that the anisotropy of VSR has a cosmological origin, perhaps a primordial magnetic field. Such fields $B
$ have been bounded by $10^{-17} G<B<10^{-9} G$ \cite{durrer}. Assuming that these primordial magnetic fields induce the electron spin flip, we get an estimation:

\begin{eqnarray}
|E_{\uparrow} - E_{\downarrow}| &=& \frac{\left|m_{L}^{2} - m_{R}^{2}\right|}{2M} = g\mu_{B} B \precsim 10^{-17} \textrm{eV} \nonumber \\
\rightarrow \left|m_{L}^{2} - m_{R}^{2}\right| & \precsim & 10^{-11} \textrm{eV}^{2}.
\end{eqnarray}

This bound is very strong. This means that $m_L = m_R$ is a excellent approximation probably in almost every case. However, the possibility of an Electron spin precession must not be ignored.\\

%%%%%%%%%%%%%%%%%%%%%%%%%%%%%%%%%%%%%%%%%Lepton-Gauge boson interactions%%%%%%%%%%%%%%%%%%%%%%%%%%%%%%%%%%%%%%%%%%%%%%%%%%%%%%%%%%%%%%%%%%%%%%%%%%%%%

\section{Lepton-Gauge boson interactions}
\label{Chap: Lepton-Gauge boson interactions.}

We now consider three lepton families $e_{b}$, $\nu_{b}$, $b=1 \ldots 3$. Keeping to first order in the gauge fields because higher terms are strongly suppressed by the smallness of the gauge coupling and the mass terms introduced by VSR, the interaction terms in Fourier space are:

\begin{eqnarray}
\label{L lept gauge}
\mathcal{L}^{int}_{lept,gauge} &=& - \frac{gg'}{\sqrt{g^2+g'^2}}\bar{e}_{b}\left(\delta^{ba}\gamma^{\mu} + \frac{\slashed{n}}{2}[\bar{m}^2]^{ba}(n\cdot (k+q))^{-1}(n\cdot q)^{-1}n^{\mu}\right)e_{a}A_{\mu} \nonumber \\
&& - \frac{g'^2}{\sqrt{g^2+g'^2}}\bar{e}_{b}\left(\delta^{ba}\gamma^{\mu} + \frac{\slashed{n}}{2}[\bar{m}^2]^{ba}(n\cdot (k+q))^{-1}(n\cdot q)^{-1}n^{\mu}\right)e_{a}Z_{\mu} \nonumber \\
&& + \frac{\sqrt{g^2+g'^2}}{2}\bar{e}_{bL}\left(\delta^{ba}\gamma^{\mu} + \frac{\slashed{n}}{2}[\bar{m}_L^2]^{ba}(n\cdot (k+q))^{-1}(n\cdot q)^{-1}n^{\mu}\right)e_{aL}Z_{\mu} \nonumber \\
&& - \frac{\sqrt{g^2+g'^2}}{2}\bar{\nu}_{bL}\left(\delta^{ba}\gamma^{\mu} + \frac{\slashed{n}}{2}[\bar{m}_L^2]^{ba}(n\cdot (k+q))^{-1}(n\cdot q)^{-1}n^{\mu}\right)\nu_{aL}Z_{\mu} \nonumber \\
&& - \frac{g}{\sqrt{2}}\bar{\nu}_{bL}\left(\delta^{ba}\gamma^{\mu} + \frac{\slashed{n}}{2}[\bar{m}_L^2]^{ba}(n\cdot (k+q))^{-1}(n\cdot q)^{-1}n^{\mu}\right)e_{aL}W^{+}_{\mu} \nonumber \\
&& - \frac{g}{\sqrt{2}}\bar{e}_{bL}\left(\delta^{ba}\gamma^{\mu} + \frac{\slashed{n}}{2}[\bar{m}_L^2]^{ba}(n\cdot (k+q))^{-1}(n\cdot q)^{-1}n^{\mu}\right)\nu_{aL}W^{-}_{\mu},
\end{eqnarray}

where $k$ and $q$ are the momentum of gauge and lepton fields respectively, we have used $u_{R,L} = \frac{1 \pm \gamma^{5}}{2}$ and:

\begin{eqnarray}
\bar{m}^2 = \frac{(1-\gamma^5)\bar{m}^2_{L} + (1+\gamma^5)\bar{m}_R^2}{2},
\end{eqnarray}

where $\bar{m}^{2}_{L}$ and $\bar{m}^{2}_{R}$ are $3 \times 3$ non diagonal hermitian matrices.\\

To zero order on $\bar{m}^{2}$, from (\ref{L lept gauge}), we can see that $e_{L}$, $e_{R}$ and $\nu_{L}$ are the flavor states, and \ $e_{L}$ and
$e_{R}$ are the mass states, but $\nu_{L}$ is not a mass eigenstate because for it the leading non-zero mass is $\bar{m}$.\\

Then, for neutrinos, the relation between both states is:

\begin{eqnarray}
\label{nuLM}
\nu^{M}_{L} = V_{l} \nu_{L},
\end{eqnarray}

where $\nu^{M}_{L}$ is the mass state and $V_{l}$ is a unitary transformation, such that $V_{l}\bar{m}_{L}^{2}V_{l}^{\dag}$ is a diagonal matrix. Because the mass and flavor states of the neutrinos are not the same, we will have an oscillation between different states, where $V_{l}$ is the mixing matrix, that correspond to the Pontecorvo-Maki-Nakagawa-Sakata (PMNS) matrix:

\begin{eqnarray}
\label{Mix Matrix}
V_l^{\dag} = \left(
        \begin{array}{ccc}
          c_{13}c_{12} & c_{13}s_{12} & e^{-i\delta}s_{13} \\
          -s_{23}s_{13}c_{12}e^{i\delta}-c_{23}s_{12} & -s_{23}s_{13}s_{12}e^{i\delta}+c_{23}c_{12} & s_{23}c_{13} \\
          -c_{23}s_{13}c_{12}e^{i\delta}+s_{23}s_{12} & -c_{23}s_{13}s_{12}e^{i\delta}-s_{23}c_{12} & c_{23}c_{13} \\
        \end{array}
      \right),
\end{eqnarray}

with $c_{ij} = \cos(\theta_{ij})$ and $s_{ij} = \sin(\theta_{ij})$. Therefore, a VSR non-diagonal mass matrix term is a natural form to describe neutrino oscillations.\\

On the other side, if we include the terms with VSR mass in (\ref{L lept gauge}), we can see that we need a \textit{non-local unitary} transformation to diagonalize the interactions and obtain the flavor states. Actually, this means that we will have oscillation in all leptons. However, to
introduce this unitary transformation is complicated and unnatural, so we will think of these terms as new very small interactions of the order of VSR
parameters. These interactions will generate transition from one family to another. This is the subject of the next section.\\

%%%%%%%%%%%%%%%%%%%%%%%%%%%%%%%%%%%%%%%%%X \rightarrow Y+ \gamma%%%%%%%%%%%%%%%%%%%%%%%%%%%%%%%%%%%%%%%%%%%%%%%%%%%%%%%%%%%%%%%%%%%%%%%%%%%%%

\section{$X \rightarrow Y+ \gamma$}
\label{Chap: X Y + gamma}

One of the flavour changing \ interactions is:

\begin{eqnarray}
-\frac{e}{2}\bar{e}_{m}[\bar{m}^2]^{mn}\slashed{n}(n\cdot (k+q))^{-1}(n\cdot A)(n\cdot q)^{-1}e_{n}, \nonumber
\end{eqnarray}

where $e$ is the electron charge. So, we have the process given by \textbf{Figure 1}, $X \rightarrow Y + \gamma$. If $X$ and $Y$ are leptons with $m_{X} > m_{Y}$, the corresponding decay rate is given by:

\begin{figure}
\caption{}
\begin{center}
\includegraphics[scale=0.6]{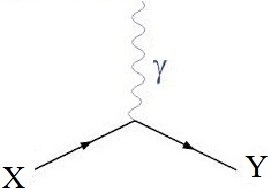}
\end{center}
\end{figure}

\begin{eqnarray}
d\Gamma = 4\pi\alpha \left(\left|[\bar{m}_L^2]_{(XY)}\right|^2+\left|[\bar{m}_R^2]_{(XY)}\right|^2\right)\frac{\left(n_i\epsilon_{pi}(k)\right)\left(n_j\epsilon^{*}_{pj}(k)\right)}{\left(n_l p_l\right)\left(n_m q_m\right)} \frac{(2\pi)^4\delta^{(4)}(q-p-k)}{2 m_X}\frac{d^3k}{(2\pi)^32E_{\gamma}(k)} \frac{d^3p}{(2\pi)^32E_Y(p)}, \nonumber
\end{eqnarray}

where $\alpha = \frac{e^2}{4\pi} \simeq \frac{1}{137}$ is the fine-structure constant, $E_{\gamma}(k) = k$, $E_Y(p) = \sqrt{p^2+m_Y^2}$, $\epsilon_{pi}$ is the photon polarization and $n_i$ is the space component of the null-vector $n^{\mu}$. We can choose $|\vec{n}| = 1$. Because of $\vec{n}$, this decay have a privileged direction, given by the polarization. We will study the unpolarized case, so we must sum on $p$ and use:

\begin{eqnarray}
\sum_p \epsilon_{pi}(k)\epsilon^{*}_{pj}(k) = \delta_{ij} - \frac{k_i k_j}{|\vec{k}|}. \nonumber
\end{eqnarray}

Now, if we evaluate $\Gamma$ on the $X$ particle rest frame, such that $q=(m_{X},\vec{0})$ with $m_{X} \gg m_{Y}$, and we use $\vec{n} = \hat{z}$, we
obtain that:

\begin{eqnarray}
\label{Decay Rate}
\Gamma(X \rightarrow Y + \gamma) \simeq \frac{\alpha\left(\left|[\bar{m}_L^2]_{(XY)}\right|^2+\left|[\bar{m}_R^2]_{(XY)}\right|^2\right)}{4 m^3_X}.
\end{eqnarray}

On the other side, we have the known process $X \rightarrow Y + \bar{\nu}_{Y} + \nu_{X}$ (See \textbf{Figure 2}), where the decay rate is:

\begin{figure}
\caption{}
\begin{center}
\includegraphics[scale=0.6]{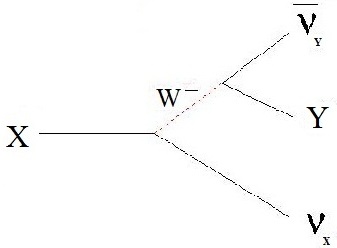}
\end{center}
\end{figure}

\begin{eqnarray}
\label{Decay Rate 0}
\Gamma(X \rightarrow Y + \bar{\nu}_Y + \nu_X) \simeq \frac{G_F^2m_X^5}{192 \pi^3},
\end{eqnarray}

with $G_{F} =1.01 \times 10^{-5} m_{P}^{-2}$ and $m_{P}$ is the proton mass.\\

So, we can compute the branching ratio given by:

\begin{eqnarray}
\label{B}
B(X,Y) &=& \frac{\Gamma(X \rightarrow Y + \gamma)}{\Gamma(X \rightarrow Y + \bar{\nu}_Y + \nu_X)} \nonumber \\
&=& \frac{48\pi^3\alpha\left(\left|[\bar{m}_L^2]_{(XY)}\right|^2+\left|[\bar{m}_R^2]_{(XY)}\right|^2\right)}{m^8_X G_F^2} \nonumber \\
&=& 47.05 \times 10^{10} \pi^3 \alpha \left(\frac{m_P}{m_X}\right)^4 \left(\left|\frac{[\bar{m}_L^2]_{(XY)}}{m_X^2}\right|^2+\left|\frac{[\bar{m}_R^2]_{(XY)}}{m_X^2}\right|^2\right).
\end{eqnarray}

Using (\ref{Mix Matrix}), we can see that:

\begin{eqnarray}
\label{mLR mod}
\left|[\bar{m}_L^2]_{(e\mu)}\right|^2 &=& c_{13}^2\left(\left(\frac{s_{23}s_{13}(c_{12}^2-s_{12}^2)}{2}+s_{12}c_{23}c_{12}\cos(\delta)\right)\delta m^2 + s_{23}s_{13}\Delta m^2\right)^2 \nonumber \\ &&+ c_{23}^2c_{13}^2s_{12}^2c_{12}^2\sin^2(\delta)\left(\delta m^2\right)^2,
\end{eqnarray}

where $\delta m^2 = m_2^2-m_1^2$ and $\Delta m^2 = m_3^2 - \frac{m_1^2+m_2^2}{2}$, with $m_{i}$ the neutrino masses corresponding to different families. In this model, the mixing angles have the usual values. That is \cite{Fogli}:

\begin{eqnarray}
s_{12}^2 &=& 0.307 \nonumber \\
s_{13}^2 &=& \left\{\begin{array}{c}
               0.0241\textrm{ (NH)} \\
               0.0244\textrm{ (IH)}
             \end{array}\right. \nonumber \\
s_{23}^2 &=& \left\{\begin{array}{c}
               0.386\textrm{ (NH)} \\
               0.392\textrm{ (IH)}
             \end{array}\right. \nonumber \\
\delta &=& \left\{\begin{array}{c}
               1.08\pi\textrm{ (NH)} \\
               1.09\pi\textrm{ (IH)}
             \end{array}\right. \nonumber \\
\delta m^2 &=& 7.54 \times 10^{-5}\textrm{ }[eV^2] \nonumber \\
\Delta m^2 &=& \left\{\begin{array}{c}
               2.43 \times 10^{-3}\textrm{ }[eV^2]\textrm{ (NH)} \\
               2.42 \times 10^{-3}\textrm{ }[eV^2]\textrm{ (IH)}
             \end{array}\right., \nonumber
\end{eqnarray}

where (NH) is normal hierarchy and (IH) inverted hierarchy. So, if we evaluate it in (\ref{mLR mod}), we obtain:

\begin{eqnarray}
\label{mL value}
\left|[\bar{m}_L^2]_{(e\mu)}\right| = \left\{\begin{array}{c}
                                               2.07 \times 10^{-4}\textrm{ }[eV^2]\textrm{ (NH)} \\
                                               2.10 \times 10^{-4}\textrm{ }[eV^2]\textrm{ (IH)}
                                             \end{array}\right.
\end{eqnarray}

and the Branching ratio with (\ref{mL value}) is:

\begin{eqnarray}
\label{L quark 0}
B = \left\{\begin{array}{c}
            4.43 \times 10^{-25}\textrm{ }[eV^2]\textrm{ (NH)} \\
            4.56 \times 10^{-25}\textrm{ }[eV^2]\textrm{ (IH)}
           \end{array}\right.. \nonumber
\end{eqnarray}

The best upper limit to the branching ratio is $B(\mu,e) <5.7 \times 10^{-13}$ \cite{Adam}. So the predicted branching ratio is much smaller than the current experimental bound.\\

%%%%%%%%%%%%%%%%%%%%%%%%%%%%%%%%%%%%%%%%%Canonical Quantization%%%%%%%%%%%%%%%%%%%%%%%%%%%%%%%%%%%%%%%%%%%%%%%%%%%%%%%%%%%%%%%%%%%%%%%%%%%%%

\section{Canonical Quantization}
\label{Chap: Canonical Quantization.}

Since VSR introduce non local terms, it is quite important to show how canonical quantization works in this case. Moreover, we have to check that it gives the same results as the path integral quantization.\\

Since non-locality means that the equations of motion are integral-differential equations, we have to fix a point of view about the quantization of such theories. Our perspective is to accept the results of the path integral quantization, which is a lagrangian quantization and better defined in this case. However, to understand the particle contain of the model, we must have a canonical formulation in terms of creation and annihilation operators. With this in mind, we will develop a canonical second quantization, such that it agrees with the path integral quantization, although we will have to introduce anticommutation rules that are non-canonical. We believe that this is due to the presence of second class constraints, and thus it is the Dirac bracket that define the anticommutation relations instead of the Poisson bracket. Aside from this subtle point that needs to be clarified in the future, we obtain a canonical second quantization which permits a particle interpretation of the model. The propagator defined in the canonical formulation coincides with the path
integral result. The creation operators describes particles of definite momentum and charge.\\

We start from a local leptonic lagrangian \cite{nonlocal}, where the canonical formalism is well defined:

\begin{eqnarray}
\label{Lagran}
\mathcal{L} = i\bar{\psi}\slashed{\partial}\psi - M\bar{\psi}\psi + i\bar{\chi}(n\cdot\partial)\phi + i\bar{\phi}(n\cdot\partial)\chi + \frac{im}{2}\bar{\chi}\slashed{n}\psi - \frac{im}{2}\bar{\psi}\slashed{n}\chi + \frac{im}{2}\bar{\phi}\psi - \frac{im}{2}\bar{\psi}\phi,
\end{eqnarray}

where $\psi$ is the lepton field, $\phi$ and $\chi$ are auxiliary fields and $n_{\mu} = (n_0,n_{i})$, so that $\mid \vec{n} \mid =1$ and $n^{\mu} = (n_0,-n^{i} )$. The lagrangian equations of motion are:

\begin{eqnarray}
\label{eqs 0}
i\slashed{\partial}\psi - M\psi - \frac{im}{2}\slashed{n}\chi - \frac{im}{2}\phi &=& 0 \nonumber \\
(n\cdot\partial)\phi + \frac{m}{2}\slashed{n}\psi &=& 0 \nonumber \\
(n\cdot\partial)\chi + \frac{m}{2}\psi &=& 0.
\end{eqnarray}

from which we can deduce:

\begin{eqnarray}
\label{Aux Fields}
\phi &=& - \frac{m}{2} (n \cdot \partial)^{-1} \slashed{n} \psi  \\
\chi &=& - \frac{m}{2} (n \cdot \partial)^{-1} \psi \nonumber\\
\bar{\phi} &=& - \frac{m}{2} (n \cdot \partial)^{-1}  \bar{\psi} \slashed{n} \nonumber\\
\bar{\chi} &=& - \frac{m}{2} (n \cdot \partial)^{-1}  \bar{\psi} \nonumber
\end{eqnarray}

and the equation of motion given by:

\begin{eqnarray}
\label{Leptonic Eq}
\left(i\left(\slashed{\partial} + \frac{m^2}{2}\slashed{n}(n\cdot \partial)^{-1}\right) - M\right)\psi = 0.
\end{eqnarray}

Now, the canonical conjugated variables are:

\begin{eqnarray}
\label{Canon Var}
P_{\psi} &=& i \psi^{\dag} \nonumber \\
P_{\chi} &=& i n_0\bar{\phi} \nonumber \\
P_{\phi} &=& i n_0\bar{\chi}
\end{eqnarray}

and the local Hamiltonian is given by:

\begin{eqnarray}
\label{Hamilt L}
\mathcal{H}_{L} = - P_{\psi}\gamma^0\left(\gamma^i\partial_i\psi + iM\psi - \frac{m}{2}\slashed{n}\chi - \frac{m}{2}\phi\right) + \frac{P_{\phi}}{n_0}\left(n^i\partial_i\phi - \frac{m}{2}\slashed{n}\psi\right) + \frac{P_{\chi}}{n_0}\left(n^i\partial_i\chi - \frac{m}{2}\psi\right).
\end{eqnarray}

Using the canonical commutation relations:

\begin{eqnarray}
\label{pb}
\{\psi(x)_{i},P_{\psi}(x')_{j}\}_{\textrm{eqt}} &=& i\delta(\vec{x} - \vec{x}')\delta_{i j} \\
\{\phi(x)_{i},P_{\phi}(x')_{j}\}_{\textrm{eqt}} &=& i\delta(\vec{x} - \vec{x}')\delta_{i j} \nonumber\\
\{\chi(x)_{i},P_{\chi}(x')_{j}\}_{\textrm{eqt}} &=& i\delta(\vec{x} - \vec{x}')\delta_{i j} \nonumber
\end{eqnarray}

and (\ref{Hamilt L}) we reproduce the lagrangian equations of motion (\ref{eqs 0}).\\

\subsection{Second Quantization}

The most general solution of equation (\ref{Leptonic Eq}) is:

\begin{eqnarray}
\label{Sol Dirac}
\psi(t,\vec{x}) \rightarrow \hat{\psi}(t,\vec{x}) &=& \sum_{s} \left(\hat{\psi}^{+}_{(s)}(t,\vec{x}) + \hat{\psi}^{-}_{(s)}(t,\vec{x})\right) \\
\label{Sol pm}
\hat{\psi}^{+}_{(s)}(t,\vec{x}) &=& \int \frac{d^3k}{(2\pi)^3} \frac{1}{\sqrt{2E(\mathbf{k})}} \hat{a}_s(\vec{k}) u_{(s)}(\vec{k})e^{-iE(\mathbf{k})t + i\vec{k}\cdot\vec{x}} \nonumber \\
\hat{\psi}^{-}_{(s)}(t,\vec{x}) &=& \int \frac{d^3k}{(2\pi)^3} \frac{1}{\sqrt{2E(\mathbf{k})}} \hat{b}^{\dag}_s(\vec{k}) v_{(s)}(\vec{k})e^{iE(\mathbf{k})t - i\vec{k}\cdot\vec{x}},
\end{eqnarray}

where $\hat{a}^{\dag}$ and $\hat{b}^{\dag}$ are the creation operators for particles and antiparticles respectively and $u_{(s)}(\vec{k})$ and $v_{(s)}(\vec{k})$ are solutions of (see \textbf{Appendix C}).

\begin{eqnarray*}
\left(\slashed{k} - \frac{m^2}{2}\slashed{n}(n \cdot k)^{-1} - M \right)u_{(s)} &=& 0 \\
\left(\slashed{k} - \frac{m^2}{2}\slashed{n}(n \cdot k)^{-1} + M \right)v_{(s)} &=& 0
\end{eqnarray*}

Using the equations of motion, we get:

\begin{eqnarray}
\label{HL}
H_{L} &=& i\int d^3x \left(- \psi^{\dag}\gamma^0\left(\gamma^i\partial_i\psi + iM\psi - \frac{m}{2}\slashed{n}\gamma^0\chi - \frac{m}{2}\gamma^0\phi\right) + \bar{\chi}\left(n^i\partial_i\phi - \frac{m}{2}\gamma^0\slashed{n}\psi\right) \right. \nonumber \\
&& \left.  + \bar{\phi}\left(n^i\partial_i\chi - \frac{m}{2}\gamma^0\psi\right)\right) \nonumber \\
&=& i\int d^3x \left(\psi^{\dag}\gamma^0\left(\gamma^0\partial_0\psi\right) + \bar{\chi}\left(n_0\partial_0\phi\right) + \bar{\phi}\left(n_0\partial_0\chi\right)\right) \nonumber \\
&=& i\int d^3x \left(\psi^{\dag}\partial_0\psi + \frac{n_0m^2}{2}\left((n\cdot\partial)^{-1}\psi^{\dag}\right)\gamma^0\slashed{n}\left((n\cdot\partial)^{-1}\partial_0\psi\right)\right).
\end{eqnarray}

From (\ref{HL}), we see that $\psi$ and $\psi^{\dag}$ are no longer canonically conjugated variables as in (\ref{pb}). Instead they satisfy the anticommutation relations given by (\ref{canonical}).\\

Finally, if we use the properties given by (\ref{ortho}), we obtain:

\begin{eqnarray}
\label{H Operator}
H_{L} &=& \sum_{s} \int \frac{d^3k}{(2\pi)^3} E(\mathbf{k})\left(\hat{a}^{\dag}_s(\vec{k})\hat{a}_s(\vec{k}) - \hat{b}_s(\vec{k})\hat{b}^{\dag}_s(\vec{k})\right) \nonumber \\
&=& \sum_{s} \int \frac{d^3k}{(2\pi)^3} E(\mathbf{k})\left(\hat{a}^{\dag}_s(\vec{k})\hat{a}_s(\vec{k}) + \hat{b}^{\dag}_s(\vec{k})\hat{b}_s(\vec{k})\right).
\end{eqnarray}

Besides, we can prove that the electric charge is:

\begin{eqnarray}
\label{charge}
Q = \int d^3x \left(\psi^{\dag}\psi + \frac{n_0m^2}{2}\left((n\cdot \partial)^{-1}\psi^{\dag}\right)\gamma^0\slashed{n}\left((n\cdot \partial)^{-1}\psi\right)\right).
\end{eqnarray}

Notice that in (\ref{charge}) the same combination of $\psi$ and $\psi^{\dag}$ appears compared to (\ref{HL}). This must be so, because $Q$ will generate $U(1)$ gauge transformation on the field upon using the anticommutation relations. We can use (\ref{Sol Dirac}) to obtain:

\begin{eqnarray}
\label{Q Operator}
Q &=& \sum_{s} \int \frac{d^3k}{(2\pi)^3} \left(\hat{a}^{\dag}_{s}(\vec{k})\hat{a}_s(\vec{k}) + \hat{b}_{s}(\vec{k})\hat{b}^{\dag}_s(\vec{k})\right) \nonumber \\
&=& \sum_{s} \int \frac{d^3k}{(2\pi)^3} \left(\hat{a}^{\dag}_{s}(\vec{k})\hat{a}_s(\vec{k}) - \hat{b}^{\dag}_s(\vec{k})\hat{b}_{s}(\vec{k})\right).
\end{eqnarray}

Now, the energy momentum tensor is:

\begin{eqnarray}
\label{Mom Tensor}
T_{\mu \nu} = i\bar{\psi}\gamma_{\nu}\partial_{\mu}\psi - \eta_{\mu \nu}\bar{\psi}\left(i\slashed{\partial}+\frac{im^2}{2}\slashed{n}(n\cdot \partial)^{-1}-M\right)\psi + \frac{im^2}{2}\left((n\cdot \partial)^{-1}\bar{\psi}\right)\slashed{n}n_{\nu}\partial_{\mu}\left((n\cdot \partial)^{-1}\psi\right),
\end{eqnarray}

where $\gamma_{\mu} = (\gamma^{0},-\gamma^{i})$. Then, the momentum operator is:

\begin{eqnarray}
\label{Mom}
P^i &=& \int d^3x T^{i 0} \nonumber \\
&=& \int d^3x \left(- i\bar{\psi}\gamma^0\partial_i\psi - \frac{im^2}{2}\left((n\cdot \partial)\bar{\psi}\right)\slashed{n}n_0\partial_i\left((n\cdot \partial)\psi\right)\right) \nonumber \\
&=& - i \int d^3x \left(\psi^{\dag}\partial_i\psi + \frac{n_0m^2}{2}\left((n\cdot \partial)^{-1}\psi^{\dag}\right)\gamma^0\slashed{n}\left((n\cdot \partial)^{-1}\partial_i\psi\right)\right),
\end{eqnarray}

where we used that $\partial^{\mu} = (\partial_{0},-\partial_{i})$. Then, following the same procedure to calculate (\ref{H Operator}) and (\ref{Q
Operator}), we obtain:

\begin{eqnarray}
P^i &=& \sum_{s} \int \frac{d^3k}{(2\pi)^3} k^i\left(\hat{a}^{\dag}_{s}(\vec{k})\hat{a}_s(\vec{k}) - \hat{b}_{s}(\vec{k})\hat{b}^{\dag}_s(\vec{k})\right) \nonumber \\
&=& \sum_{s} \int \frac{d^3k}{(2\pi)^3} k^i\left(\hat{a}^{\dag}_{s}(\vec{k})\hat{a}_s(\vec{k}) + \hat{b}^{\dag}_s(\vec{k})\hat{b}_{s}(\vec{k})\right).
\end{eqnarray}

\subsection{Propagator}

Assuming the standard anticommutation relations for the creation and annihilation operators:

\begin{eqnarray}
\label{creation}
  \begin{array}{lll}
    \{ a_{s} ( p ) ,a_{r} ( q )^{\uparrow} \} = ( 2 \pi )^{3} \delta^{( 3 )} (
    p-q ) &  & \begin{array}{lll}
      \{ b_{s} ( p ) ,b_{r} ( q )^{\uparrow} \} = ( 2 \pi )^{3} \delta^{( 3 )}
      ( p-q ), &  &
    \end{array}
  \end{array}
\end{eqnarray}

we compute:

\begin{eqnarray*}
<0|\psi_{a}(x)\bar{\psi}_{b}(y)|0> &=& \int \frac{d^3p}{(2\pi)^{3}} \frac{1}{2E_{p}} \sum_{s} u_{sa}(p)\bar{u}_{sb}(p) e^{-ip\cdot (x-y)} \\
&=& \int \frac{d^3p}{(2\pi)^{3}} \frac{1}{2E_{p}} \left(\left(\slashed{p} + M - \frac{m^{2}}{2}\frac{\slashed{n}}{(n\cdot p)}\right)_{a b} e^{-ip\cdot (x-y)}\right) \\
&=& \left(i\slashed{\partial}_{x} + M + i\frac{m^{2}}{2}\frac{\slashed{n}}{(n\cdot\partial_{x})} \right)_{a b} \int \frac{d^3p}{(2\pi)^{3}}\frac{1}{2E_{p}}e^{-ip\cdot (x-y)}
\end{eqnarray*}

and:

\begin{eqnarray*}
<0|\bar{\psi}_{b}(y)\psi_{a}(x)|0> &=& \int \frac{d^3p}{(2\pi)^{3}} \frac{1}{2E_{p}} \sum_{s} \bar{v}_{sb}(p) v_{sa}(p) e^{ip\cdot (x-y)} \\
&=& \int \frac{d^3p}{(2\pi)^{3}} \frac{1}{2E_{p}} \left(\left(\slashed{p} - M - \frac{m^{2}}{2}\frac{\slashed{n}}{(n\cdot p)}\right)_{a b} e^{ip\cdot (x-y)}\right) \\
&=& - \left(i\slashed{\partial}_{x} + M + i\frac{m^{2}}{2}\frac{\slashed{n}}{(n\cdot\partial_{x})} \right)_{a b} \int \frac{d^3p}{(2\pi)^{3}}\frac{1}{2E_{p}}e^{ip\cdot (x-y)}.
\end{eqnarray*}

That is:

\begin{eqnarray*}
S_{F} ( x-y ) = \int \frac{d^{4} p}{(2\pi)^4} \frac{i \left(\slashed{p} + M - \frac{m^2}{2}\frac{\not{n}}{n\cdot p}\right)}{p^{2} - M^2 - m^2 + i\varepsilon} e^{- ip\cdot (x-y)}
\end{eqnarray*}

which coincides with the path integral result. This calculation shows that the relations (\ref{creation}) are correct. It follows that the model describes particles of definite energy-momentum and charge.\\

Finally, we have that the canonical anticommutation relations are, after using the equations of motion of the auxiliary fields:

\begin{eqnarray}
\label{canonical}
\{\psi_{a}(x),\psi_{b}(y)^{\dag}\}_{\textrm{eqt}} &=& \int \frac{d^{3} p}{(2\pi)^{3}} \frac{1}{2E_{p}} \sum_{s} \left(u_{sa}(p)u^{\dag}_{sb}(p)e^{-i\vec{p}(\vec{x}-\vec{y})} + v_{sa}(p)v^{\dag}_{sb}(p)e^{i\vec{p}(\vec{x}-\vec{y})}\right) \nonumber \\
&=& \int \frac{d^{3} p}{(2\pi)^{3}} \frac{1}{2E_{p}} \sum_{s} \left(u_{sa}(p)u^{\dag}_{sb}(p) + v_{sa}(-p)v^{\dag}_{sb}(-p)\right)e^{-i\vec{p}(\vec{x}-\vec{y})} \nonumber \\
&=& \int \frac{d^{3} p}{(2\pi)^3}e^{-i\vec{p}(\vec{x}-\vec{y})}\left(1-\frac{m^2}{2n_{0}\left(E_{p}^{2}-(\hat{n}\cdot\vec{p})^{2}\right)}\slashed{n}\gamma^0\right)_{a b}.
\end{eqnarray}

Just like we said before, we believe that (\ref{canonical}) came from a Dirac bracket by the presence of second class constraints on the model and it shall be clarified in the future.\\

%%%%%%%%%%%%%%%%%%%%%%%%%%%%%%%%%%%%%%%%%Conclusions%%%%%%%%%%%%%%%%%%%%%%%%%%%%%%%%%%%%%%%%%%%%%%%%%%%%%%%%%%%%%%%%%%%%%%%%%%%%%

\section*{Conclusions and Open Problems}

In this paper, we applied the VSR formalism to the Electroweak Standard Model. This modification admits the generation of a neutrino mass without lepton number violation and without sterile neutrinos or another types of additional  particles. However a non local term is necessary. So, the VSR EW SM is a simple theory with $SU(2)_L \times U(1)_R$ symmetry, with the same number of leptons and gauge fields as in the Electroweak Standard Model, but now we have non local mass terms that violate Lorentz invariance. Besides, it is renormalizable and unitary.\\

First, we review the formulation of Yang Mills fields in VSR, developed in \cite{ar1}, and then we used this to define the VSR EW SM gauge bosons and find the equations of motion after spontaneous symmetry breaking. With this, we concluded that the number of the degrees of freedom are not modified with respect to the usual Electroweak SM, $W_{\mu}$ and $Z_{\mu}$ having three degrees of freedom and $A_{\mu}$ has two. However the masses are modified; $W^{\pm}_{\mu}$ and $Z_{\mu}$ masses are $M_W = \frac{vg}{2}$ and $M_Z = \frac{v\sqrt{g^2+g'^2}}{2}$ respectively for longitudinal polarization with respect to $n_{\mu}$ and $M_W = \sqrt{\frac{v^2g^2}{4}+m^2_G}$ and $M_Z = \sqrt{\frac{v^2(g^2+g'^2)}{4}+m^2_G}$ for perpendicular polarizations with respect to $n_{\mu}$ respectively. On the other side, the photon, $A_{\mu}$, has a unique mass $M_A=m_G$ for the two polarizations. We presented some bounds on $m_G$. In a future work, all these prediction should be developed to be studied in appropriated experiments, for example at the LHC.\\

In the second place, we solved the equations of motion for the leptons. A modified dispersion relation is produced and, in the particular case of neutrinos, they obtain mass without lepton number violation or sterile neutrinos. Besides, we can produce neutrino oscillations. For the electron (muon,tau), we obtained a interesting effect in the case $m_L\neq m_R$, an Electron Spin oscillation. This means that the electrons (muon and tau) are actually composed by two different states with slightly different masses. In fact, we found a extremely strong bound. This is $\left|m_{L}^{2} - m_{R}^{2}\right| \precsim 10^{-11} \textrm{eV}^{2}$. Therefore, $m_L = m_R$ is an excellent approximation.\\

In the third place, we analyzed the leptons gauge bosons interactions to study new process forbidden in the usual Electroweak model. In particular, we computed the decay rate for $X->Y+\gamma$, where $X$ and $Y$ are leptons with $m_X>m_Y$. We obtain a more restrictive condition to the Branching ratio compared with the best experimental bounds available today.\\

Finally, we analyzed the canonical quantization of the model. For this, we used auxiliary fields to eliminate the non local terms and obtain a local hamiltonian. Then, we quantize. To come back to the non local formalism, we must use the equations of motion of the auxiliary fields. However, they are integral-differential equations. This produce an non-canonical anticommutation relation for the fermion field. So, we decided to accept the results of the path integral quantization like the correct point of view about the quantization and we proved that this produce the correct expressions of the propagator, hamiltonian and charge operator in terms of creation and annihilation operators within the canonical second quantization. We believe that this non-canonical anticommutation relation came from a Dirac bracket by the presence of second class constraints on the model. This point shall be clarified in a future work.\\

In the present work, we did not included quarks in the formalism. We leave the implementation of this part of the VSR EW SM for a future publication. Meanwhile, many interesting applications of the model open up: among them to study the processes that have been observed at the LHC, to put bounds on the parameters of the model and/or describe new Physics beyond the SM, to be ready for the precision tests that will be available at the next run of the LHC.\\

%%%%%%%%%%%%%%%%%%%%%%%%%%%%%%%%%%%%%%%%%Acknowledgements%%%%%%%%%%%%%%%%%%%%%%%%%%%%%%%%%%%%%%%%%%%%%%%%%%%%%%%%%%%%%%%%%%%%%%%%%%%%%

\section*{Acknowledgements}

The work of P. Gonz\'alez and R. Avila has been partially financed by Fondecyt 1110378 and Anillo ACT 1102. The work of P. Gonz\'alez has been partially financed by CONICYT Programa de Postdoctorado FONDECYT $N^o$ 3150398. The work of JA is partially supported by Fondecyt 1110378, Fondecyt 1150390 and Anillo ACT 1102. The authors want to thank C. Aulakh for very interesting and enlightening discussions.\\

%%%%%%%%%%%%%%%%%%%%%%%%%%%%%%%%%%%%%%%%%Bibliography%%%%%%%%%%%%%%%%%%%%%%%%%%%%%%%%%%%%%%%%%%%%%%%%%%%%%%%%%%%%%%%%%%%%%%%%%%%%%

%%%%%%%%%%%%%%%%%%%%%%%%%%%%%%%%%%%%%%%%%Appendix A%%%%%%%%%%%%%%%%%%%%%%%%%%%%%%%%%%%%%%%%%%%%%%%%%%%%%%%%%%%%%%%%%%%%%%%%%%%%%

\section*{Appendix A}

Let us study the equation (\ref{eqn}):

\begin{eqnarray}
\label{eqn2}
\left(\partial^2 + M^2\right) V_{\mu} - \left(\partial_{\mu} + m_G^2(n \cdot \partial)^{-1} n_{\mu}\right)\left((\partial \cdot V) + m_G^2(n \cdot \partial )^{-1} (n \cdot V)\right) && \nonumber \\
+ m_G^2 n_{\mu}(n \cdot \partial )^{-2}\left(\partial^2 + m_G^2\right)(n \cdot V) &=& 0.
\end{eqnarray}

If we contract with $\partial^{\mu}$ and $n^{\mu}$ respectively, we obtain:

\begin{eqnarray}
\label{defr}
\left(M^2 - m_G^2\right)(\partial \cdot V) &=& 0 \nonumber \\
\left(\partial^2 + M^2 - m_G^2\right)(n \cdot V) - (n \cdot \partial)(\partial \cdot V) &=& 0.
\end{eqnarray}

From these equations, we have two cases:\\

\textbf{I) $M = m_G$:} In this case, (\ref{eqn2}) is reduce to:

\begin{eqnarray}
\label{eqn gauge 01}
\left(\partial^2 + m_G^2\right) V_{\mu} - \partial_{\mu}(\partial \cdot V) - m_G^2(n \cdot \partial )^{-1}\partial_{\mu}(n \cdot V) &=& 0 \\
\label{eqn gauge 02}
\partial^2(n \cdot V) - (n \cdot \partial)(\partial \cdot V) &=& 0.
\end{eqnarray}

Additionally, we have a gauge invariance given by:

\begin{eqnarray}
\delta V_{\mu} &=& \partial_{\mu} \epsilon,
\end{eqnarray}

because (\ref{eqn gauge 01}) and (\ref{eqn gauge 02}) are just like (\ref{Id 1}) and (\ref{Id 2}) respectively. So we must fix the gauge. Using the Lorentz gauge $(\partial \cdot V) = 0$, we have that:

\begin{eqnarray}
\label{eqn gauge 1}
\left(\partial^2 + m_G^2\right)V_{\mu} - m_G^2(n \cdot \partial )^{-1}\partial_{\mu}(n \cdot V) &=& 0 \\
\label{eqn gauge 2}
\partial^2(n \cdot V) &=& 0,
\end{eqnarray}

but a gauge degrees of freedom survive, given by:

\begin{eqnarray}
V'_{\mu} = V_{\mu} + \partial_{\mu}\lambda \rightarrow \partial^2 \lambda = 0.
\end{eqnarray}

From (\ref{eqn gauge 2}) we see that this freedom can be used to fix $(n \cdot V) = 0$, then:

\begin{eqnarray}
\label{eqn gauge 3}
\left(\partial^2 + m_G^2\right)V_{\mu} &=& 0.
\end{eqnarray}

We use a plane wave solution to count the degrees of freedom:

\begin{eqnarray*}
V_{\mu} = \varepsilon_{\mu} e^{-i k\cdot x}, k^2 - m_G^2 &=& 0 \\
\rightarrow (k \cdot \varepsilon) = 0 \textrm{ and } (n \cdot \varepsilon) &=& 0
\end{eqnarray*}

So, the gauge field has mass $m_G$ and 2 independent polarizations (2 degrees of freedom).\\

\textbf{II) $M \neq m_G$:} In this case, (\ref{eqn2}) is reduce to:

\begin{eqnarray}
\label{eqn no proca}
\left(\partial^2 + M^2\right) V_{\mu} - m_G^2(n \cdot \partial )^{-1}\left(\partial_{\mu} - n_{\mu}(n \cdot \partial )^{-1}\partial^2\right)(n \cdot V) &=& 0 \\
(\partial \cdot V) &=& 0 \nonumber \\
\left(\partial^2 + M^2 - m_G^2\right)(n \cdot V) &=& 0. \nonumber
\end{eqnarray}

We can see that (\ref{eqn no proca}) is not a proca-like equation and it is not a gauge invariant, therefore this case is not included on \cite{Pol}. Using a plane wave solution, $V_{\mu} = \varepsilon_{\mu} e^{-i k\cdot x}$, we obtain:

\begin{eqnarray}
\left(k^2 - M^2\right)\varepsilon_{\mu} + m_G^2(n \cdot k)^{-1}\left(k_{\mu} - n_{\mu}(n \cdot k)^{-1}k^2\right)(n \cdot \varepsilon) &=& 0 \nonumber \\
(k \cdot \varepsilon) &=& 0 \nonumber \\
\left(k^2 - M^2 + m_G^2\right)(n \cdot \varepsilon) &=& 0. \nonumber
\end{eqnarray}

The third equation say us:

\begin{eqnarray}
(n \cdot \varepsilon) = \lambda \delta\left(k^2 - M^2 + m_G^2\right), \nonumber
\end{eqnarray}

where $\lambda$ is an arbitrary scalar. So, evaluating this in the others equations, we conclude that the most general solution is:

\begin{eqnarray}
\label{sol no proca}
\varepsilon_{\mu} = \Lambda_{\mu}\delta\left(k^2 - M^2\right) + \lambda(n \cdot k)^{-1}\left(k_{\mu} - (M^2-m_G^2)n_{\mu}(n \cdot k)^{-1}\right)\delta\left(k^2 - M^2 + m_G^2\right),
\end{eqnarray}

where $\Lambda_{\mu}$ is an arbitrary vector such that $(k\cdot \Lambda) = (n\cdot \Lambda) = 0$. In conclusion of this result we can say that we have three degrees of freedom: $\lambda$ (1) and $\Lambda_{\mu}$ (2). However, the mass change for different polarization. Respectively, the masses are $M^2-m_G^2$ and $M^2$.\\

%%%%%%%%%%%%%%%%%%%%%%%%%%%%%%%%%%%%%%%%%Appendix B%%%%%%%%%%%%%%%%%%%%%%%%%%%%%%%%%%%%%%%%%%%%%%%%%%%%%%%%%%%%%%%%%%%%%%%%%%%%%

\section*{Appendix B}

When $m_{L} \neq m_{R}$, the solutions can be written as states with spin in the $\hat{n}$ direction. Actually, we can prove that (\ref{u1}) and (\ref{u2}) represent the spin up and down respectively. The same thing for (\ref{u1}) and (\ref{u2}). We have that:

\begin{eqnarray}
\label{Us}
\mathcal{U}_1 = N_1
\left(
  \begin{array}{c}
                1           \\
   \frac{n_1+in_2}{n_0+n_3} \\
                1           \\
   \frac{n_1+in_2}{n_0+n_3} \\
  \end{array}
\right)\textrm{, }
\mathcal{U}_2 = N_2
\left(
  \begin{array}{c}
  -\frac{n_1-in_2}{n_0+n_3} \\
                1           \\
   \frac{n_1-in_2}{n_0+n_3} \\
               -1           \\
  \end{array}
\right)
\end{eqnarray}

in the Dirac representation, where $N_1$ and $N_2$ are normalization parameters.\\

%%%%%%%%%%%%%%%%%%%%%%%%%%%%%%%%%%%%%%%%%Appendix C%%%%%%%%%%%%%%%%%%%%%%%%%%%%%%%%%%%%%%%%%%%%%%%%%%%%%%%%%%%%%%%%%%%%%%%%%%%%%

\section*{Appendix C}

When $m_R = m_L = m$, the solutions to the VSR Dirac equation can be written:

\begin{eqnarray*}
u_{s} &=& \frac{1}{\sqrt{k_{0} + M - \frac{m^{2}}{2(n\cdot k)}n_{0}}}\left(\slashed{k} - \frac{m^{2}}{2(n \cdot k)}\slashed{n} + M\right)
\left(\begin{array}{c}
   \varphi_{s}\\
    0
\end{array}\right) \textrm{, with: }
\varphi_{1} =
\left(\begin{array}{c}
    1\\
    0
\end{array}\right) \textrm{ and }
\varphi_{2} =
\left(\begin{array}{c}
    0\\
    1
\end{array}\right) \\
v_{s} &=& \frac{1}{\sqrt{k_{0} + M - \frac{m^{2}}{2(n\cdot k)}n_{0}}}\left(\slashed{k} - \frac{m^{2}}{2(n \cdot k)}\slashed{n} - M\right)
\left(\begin{array}{c}
      0  \\
   \chi_{s}
\end{array}\right) \textrm{, with: }
\chi_{1} =
\left(\begin{array}{c}
    0\\
    1
\end{array}\right) \textrm{ and }
\varphi_{2} =
\left(\begin{array}{c}
    -1\\
     0
\end{array}\right)
\end{eqnarray*}

These solutions reduce to the standard Dirac solutions in the Pauli-Dirac representation, for $m=0$ \cite{Langacker}. They satisfy the Completeness Relations:

\begin{eqnarray}
\label{Compl Rel u}
\sum_s u_s(\vec{k})\bar{u}_s(\vec{k}) &=& \slashed{k} - \frac{m^2\slashed{n}}{2(n\cdot k)} + M \\
\label{Compl Rel v}
\sum_s v_s(\vec{k})\bar{v}_s(\vec{k}) &=& \slashed{k} - \frac{m^2\slashed{n}}{2(n\cdot k)} - M.
\end{eqnarray}

and the Orthogonality Rules given by:

\begin{eqnarray}
\label{ortho}
&& u^{\dag}_{(s)}(\vec{k})u_{(s')}(\vec{k}) = \left(2E(\mathbf{k})-\frac{n_0m^2}{(n\cdot k)}\right)\delta_{ss'} \nonumber \\
&& v^{\dag}_{(s)}(\vec{k})v_{(s')}(\vec{k}) = \left(2E(\mathbf{k})-\frac{n_0m^2}{(n\cdot k)}\right)\delta_{ss'} \nonumber \\
&& u^{\dag}_{(s)}(\vec{k})v_{(s')}(-\vec{k}) = \frac{n_0m^2}{\sqrt{n_0^2E(\mathbf{k})^2-(\vec{n}\cdot\vec{k})^2}}\delta_{ss'} \nonumber \\
&& u^{\dag}_{(s)}(\vec{k})\gamma^0\slashed{n}u_{(s')}(\vec{k}) = 2(n\cdot k)\delta_{ss'} \nonumber \\
&& v^{\dag}_{(s)}(\vec{k})\gamma^0\slashed{n}v_{(s')}(\vec{k}) = 2(n\cdot k)\delta_{ss'} \nonumber \\
&& u^{\dag}_{(s)}(\vec{k})\gamma^0\slashed{n}v_{(s')}(-\vec{k}) = 2\sqrt{n_0^2E(\mathbf{k})^2-(\vec{n}\cdot\vec{k})^2}\delta_{ss'}
\end{eqnarray}
\section*{Appendix D}
The only pole the neutrino propagator has is at $p^{2} =m_L^{2}$. In order to
see it in a simpler way , we consider the propagator derived from equation
(53) in 2 space-time dimensions.
\begin{eqnarray*}
  \gamma^{1} = \sigma_{1} =i \left( \begin{array}{c}
    0 1\\
    1 0
  \end{array} \right) & \gamma^{0} = \sigma_{3} = \left( \begin{array}{c}
    1 0\\
    0-1
  \end{array} \right) , & \alpha =- \frac{1}{2}  \frac{m_L^{2}}{n.p}\\
  \det   \left( \begin{array}{c}
    p_{0} - \alpha n_{0} i(p_{1} - \alpha n_{1} )\\
    i(p_{1} - \alpha n_{1} )-(p_{0} - \alpha n_{0} )
  \end{array} \right) = &  & \\
  -p_{0}^{2} +2 \alpha p_{0} n_{0} - \alpha^{2} n_{0}^{2} +p_{1}^{2} -2 \alpha
  p_{1} n_{1} + \alpha^{2} n_{1}^{2} = & -p^{2} +m_L^{2} & 
\end{eqnarray*}
The determinant has been computed for arbitrarily small $n.p$. We used the property that $n_\mu$ is a null vector. That is: $n^2=0$. There is the
Lorentz invariant pole only. This result holds in arbitrary space-time dimensions.
\end{document}